# The impacts of optimization algorithm and basis size on the accuracy and efficiency of variational quantum eigensolver

Xian-Hu Zha[1], Chao Zhang[2], Dengdong Fan[2], Pengxiang Xu[2], Shiyu Du [3] and Rui-Qin Zhang[4], Chen Fu[1,*]

[1]Key Laboratory of Optoelectronic Devices and Systems of Ministry of Education and Guangdong Province, College of Physics and Optoelectronic Engineering, Shenzhen University, Shenzhen 518060, China

[2]Center for Quantum computing, Peng Cheng Laboratory, Shenzhen, 518005, China

[3]Engineering Laboratory of Advanced Energy Materials, Ningbo Institute of Materials Technology and Engineering, Chinese Academy of Sciences, Ningbo, Zhejiang, 315201, China

[4]Department of Physics, City University of Hong Kong-Hong Kong SAR, China

*Email: chenfu@szu.edu.cn.

**Abstract**

Variational quantum eigensolver (VQE) is demonstrated to be the desirable methodology for quantum chemistry based on near-term quantum devices. However, several scientific issues remain to be investigated for this approach, such as the influences of optimization algorithm and basis size on the accuracy and efficiency for quantum computing. To address these issues, five molecules ($H_2$, LiH, HF, $N_2$ and $F_2$) are studied in this work using unitary coupled cluster (UCC) ansatz in second quantization description. The performance of the gradient optimization L-BFGS-B is compared with that of the direct search method COBYLA. The former converges more quickly, but the accuracy of energy surface is lower. The basis set shows a vital influence on the accuracy and efficiency. A larger basis set generally provides a more accurate energy surface, at the expense of a significant increase in computing time. The 631g basis is generally required from the energy surface of the simplest $H_2$ molecule. Based on limited quantum resources, complete active space (CAS) is suggested. With the same number of qubits, more occupied orbitals included in CAS gives a higher accuracy for the energy surface and a smaller evaluation number in the VQE optimization. Additionally, the electronic structure, such as filling fraction of orbitals, the bond strength of a molecule and the maximum nuclear charge also influences the performance of optimization, where half occupation of orbitals generally requires a large computation cost.

## 1. Introduction

With the developments of computers and quantum mechanics, computational chemistry becomes a vital research approach in material design, drug development and *etc*. In principle, both static and dynamical properties for molecules could be derived by solving the Schrödinger equation. However, on a classical computer, the exact solution can only apply to the molecule with a few atoms, because the required resource grows exponentially on the dimensionality of the system size studied.[1] Many approximate algorithms have been proposed, such as the well-known Hartree-Fock[2,3] and density functional theory (DFT).[4,5] Most methods neglect or underestimate the correlation energies, which are inaccurate in describing strongly correlated molecules.[6] However, the accuracy of the energy surface is critically important in many problems. It is required to achieve the "chemical accuracy" ($1.6\times10^{-3}$ Hartree) for predicting chemical reaction rate.[7] Efficient and accurate approaches for simulating large chemical systems are still in developing.

Motivated by the exponential complexity of simulating quantum systems, Feynman initially proposed the conjecture of universal computer to simulate physical systems in 1982.[8] Later, this conjecture was demonstrated by Lloyd.[9] Cooperated with Abrams, Lloyd also formulated that how to simulate many-body Fermi systems based on both first and second quantized descriptions.[10] Besides, they proposed a quantum algorithm known as quantum phase estimation (QPE) for finding the eigenvectors and eigenvalues of a Hamiltonian operator.[11] Based on QPE, Aspuru-Guzik *et al.* firstly performed the exact quantum chemistry calculation in polynomial time for hydrogen ($H_2$), water ($H_2O$) and hydrogen lithium (LiH) molecules.[12] However, Wecker *et al.* found that the gate-count was estimated to be $O(N^8)$ of the spinal orbitals N, based on a Trotter decomposition for QPE.[13] Although various improvements have been proposed to reduce the quantum resources,[14-17] QPE generally requires fully coherent evolution[18,19] that impedes its application in near-term quantum devices.

To reduce the coherence requirement on quantum hardware, Peruzzo and McClean *et al.* proposed a quantum-classical hybrid optimization scheme, which is known as variational quantum eigensolver (VQE).[20,21] Combined with classical optimization and state preparation based on ansatz, VQE shows a low circuit depth, and is robust to certain errors.[22] Therefore, VQE is demonstrated as the desired algorithm for the near-term quantum computers, in terms of tens of error-prone physical qubits. Combined with conventional computers, the VQE algorithm has been implemented in photonic,[20] superconducting[23,24] and trapped ion[25,26] quantum processors. According to the computational process, several terms could influence the computational efficiency and accuracy of VQE, such as the ansatz state, Hamiltonian type, optimization method and the accuracy in apparatus.[27] For the ansatz state preparation and Hamiltonian, previous schemes for reducing quantum resources based on QPE may also be utilized in VQE, such as choosing different basis sizes,[16, 17] using a truncated Taylor series instead of Trotter for simulating Hamiltonian evolution.[14] Different from QPE, the cost of VQE calculation additionally depends on the number of evaluations for optimization, and the amount of operations in state preparation and measurement.[28] Until now, the second quantization using Gaussian orbitals is the most widely used method based on the limited quantum resources available. For the ansatz states, hardware-efficient[23] and chemical inspired ansatzes[26] were proposed, but the hardware-efficient ansatz was found to be unsuitable for more than a few qubits.[29] The chemical inspired ansatz of unitary coupled cluster (UCC) presented reliable performances in physical quantum systems.[21, 25] To reduce the required qubits and the parameters in preparation of the UCC ansatz, the complete active space (CAS) approach was proposed without introducing significant loss of accuracy.[28] Regarding to the optimization method, the Nelder-Mead algorithm was determined to be robust to noises,[20] other derivative free optimization techniques were also adopted.[21] Recently, Romero *et al.* compared the performances of three derivative free optimization algorithms: Nelder-Mead,[30] Powell,[31] and COBYLA,[32] and a gradient L-BFGS-B[33] method. The results implied that COBYLA and L-BFGS-B show better performances with fewer evaluation numbers and higher energy accuracy.[28] To reduce the error in apparatus, it is critical to enhance the coherence time of qubit, and decrease the error in state preparation and reading processes. Several error mitigation approaches were proposed.[27, 34, 35]

Although many efforts have been performed to enhance the performance of VQE, many controversial issues are still existing. For instance, most reports chose the minimal basis size to elaborate the algorithm,[20, 25] but this basis is generally inaccurate in describing the energy surface.[27] CAS approach is generally required to reduce the simulation cost. How should we choose the molecular orbitals in CAS? Based on the reports of McClean *et al.*[36] and Babbush *et al.*,[37] the cost of QPE simulation is dependent on the maximum nuclear charge in a molecule and the filling fraction of orbitals. How do these electronic quantities influence on the optimization process in VQE? The COBYLA and L-BFGS-B algorithms both exhibited favorable performance in predicting the $H_4$ system.[28] What about their performances for more complex systems?

To answer these questions, the performances between the COBYLA and L-BFGS-B optimization algorithms are firstly compared in this work. L-BFGS-B converges more quickly, especially for larger molecules. Different basis sizes for five small molecules ($H_2$, $N_2$, $F_2$, LiH and HF) are studied. The 631g basis is at least required from the energy surface of the simplest $H_2$ molecule. Based on the 631g basis and the L-BFGS-B optimization algorithm, the CAS approach with varying molecular orbitals is studied. Both occupied and unoccupied molecular orbitals in the vicinity of the highest occupied orbital are simultaneously included in CAS, like that in the traditional computational chemistry. The number of function evaluation increases with a larger number of orbitals adopted. Moreover, the maximum nuclear charge, bonding energy and the filling fraction of orbitals also impact on the accuracy of energy surface and calculation efficiency. Our paper is organized as follows. Section 1 is the introduction of this work. In section 2, we briefly introduce the UCC-VQE approach, and the computational details. In section 3, the performances between the COBYLA and L-BFGS-B algorithms are compared. In section 4, the type of basis size influence on the accuracy of energy surface is discussed. For practical applications, the CAS approach with different molecular orbitals is investigated in section 5. The influence of the electronic structure on the performance of VQE is stated in section 6. In the last section, we briefly conclude our results.

## 2. Theory and computational details

The UCC-VQE approach has been discussed in previous reports.[20, 21] To solve the eigenvalue of an observable is restated as a variational problem on the Rayleigh-Ritz quotient. For the lowest eigenvalue, the corresponding eigenvector $|\Psi(\theta)>$ is that minimizes the value of $\frac{<\Psi(\theta)|H|\Psi(\theta)>}{<\Psi(\theta)|\Psi(\theta)>}$, with $H$ denotes the Hamiltonian of the system investigated. Based on the UCC ansatz, the eigenvector could be rewritten as $|\Psi(\theta)>= \exp[T(\theta)-T(\theta)^+]|\psi>$. Here, $|\psi>$ denotes a reference state, such as the Hartree-Fock state. $T(\theta)$ is the cluster operator, and $T(\theta)^+$ is the adjoint operator of $T(\theta)$. Based on this construction, the operator $\exp[T(\theta)-T(\theta)^+]$ is unitary that satisfies the requirement for quantum computation. $T(\theta)$ for an N electron system, is defined by $T(\theta)=T_1+T_2+...+T_N$. Normally, only up to the second-order terms are considered, and the corresponding approach is denoted as UCCSD. Here, $T_1=\sum_{pr}\theta_p^r a_p^+ a_r$, and $T_2=\sum_{pqrs}\theta_{pq}^{rs} a_p^+ a_q^+ a_r a_s$, with $\theta$ denoting the excitation amplitude, which is the adjustable parameter during the optimization process[25] in VQE. $a^+$ and $a$ are the Fermion creation and annihilation operators. $p$ and $q$ are the unoccupied molecular orbitals, while $r$ and $s$ denote those occupied orbitals. To map the operator $\exp[T(\theta)-T(\theta)^+]$ to quantum computer, Trotter or other expansions are generally required.

Based on the second quantization description with Gaussian orbitals, all the calculations in this work are implemented in the Qiskit code with a backend of statevector simulator.[38] The one and two electron integrals are calculated in the PySCF code.[39] The reference state is set as the Hartree-Fock state. Various basis sizes labeled as sto3g, sto6g, minao, 631g, 631g[*] and 631++g are considered for the $H_2$ molecule, while 631g is used in describing CAS approach for other molecules. The Parity encoding is adopted for mapping the Fermionic operators to qubit operators. For the UCCSD ansatz, the Trotter step is set to one. Both the COBYLA and L-BFGS-B optimization algorithms are studied, with the number of maximum iteration steps is set as 10000. The quantum exact eigensolver (QEE) is also studied for checking the performance of VQE. In order to plot the energy surface, the interatomic distance ranges from 0.5 to 3.0 Å with a step of 0.1 Å is investigated for all the molecules.

## 3. The performances between the COBYLA and L-BFGS-B algorithms

Before investigating the basis size influences on the VQE performances, the optimization algorithm is studied firstly. Previously, Romero *et al.* found that both the gradient-free COBYLA and the gradient-based L-BFGS-B were outperformed others in describing the $H_4$ molecule.[20, 28] Therefore, the performances of COBYLA and L-BFGS-B are further compared in this work. To be reliable, three molecules ($H_2$, LiH and HF) with different electrons are studied based on the minimal basis sto3g. The VQE energies from different optimization methods, the QEE energy, and the VQE evaluation numbers versus interatomic distances are provided in Figure 1. As shown in Figure 1(a), the red diamond and blue triangle show the VQE energies of $H_2$ from COBYLA and L-BFGS-B, respectively. The dark grey line presents the QEE energy. Both VQE energies coincide with that from QEE. The minimum energy is -1.136 Ha at the interatomic distance of 0.7 Å. After that, the energy increases significantly up to 2.6 Å. At the largest distance of 3.0 Å, the energy value is -0.934 Ha. The energy differences between the QEE and VQE are provided in Figure 1(b), where the lower and upper limits of y-axis are set as the chemistry accuracy of 0.0016 Ha. The tiny energy differences imply that both optimization algorithms present excellent performances in the energy surface of the simple $H_2$ molecule. Figure 1(c) shows the relationships between the evaluation numbers and interatomic distances. From the figure, L-BFGS-B shows smaller evaluation numbers compared to COBYLA. Moreover, the evaluation number of COBYLA is significantly dependent on the interatomic distance, but the value of L-BFGS-B is slightly affected by the distance. Most evaluation numbers in L-BFGS-B are determined to be 20, expect those at the distances of 1.70, 1.80 and 1.90 Å found to be 16. For COBYLA, the evaluation number is generally smaller around the equilibrium state. The minimum number is 46 at 0.70 Å. When the interatomic distance is larger than 2.0 Å, the evaluation number increases significantly, the largest value is 3537 at 2.90 Å. Based on the different evaluation numbers, the total running times for the energy curves of COYBLA and L-BFGS-B are determined to be 1211 and 922 s, respectively. The time difference is not proportional to the ratio between the evaluation numbers of different optimization algorithms, which implies that the ansatz state preparation costs much time in the computation. Figure 1(d)-1(f) present the corresponding values for the LiH molecule. From Figure 1(d), the VQE energies also seem to coincide with that of QEE, and the minimum total energy is determined to be -7.882 Ha at 1.60 Å. From the energy differences shown in Figure 1(e), the COBYLA gives a better energy accuracy compared to L-BFGS-B. The energy difference is approximate to zero between those of COBYLA and QEE. As to L-BFGS-B, a few deviations exist in the figure, but the deviations are generally smaller than the chemical accuracy in the range of interatomic distance from 0.5 to 2.7 Å. Two exceptions are determined at the distances of 2.8 and 3.0 Å, with the energy difference of 0.002 Ha. Regarding to the evaluation numbers in Figure 1(f), COBYLA shows relatively large numbers. The minimum value is 4799 at the distance of 1.0 Å. The number 10000 even occurs at some small or large distances, which number is the maximum value set previously in the optimization algorithm. The evaluation number for L-BFGS-B is constant, corresponding value is 1023. The total running times for COBYLA and L-BFGS-B are respectively $9.186 \times .0^4$ and $1.211 \times a0^4$ s. Evidently, L-BFGS-B converges more quickly in the optimization process compared to COBYLA. Figure 1(g)-1(i) are the values for the HF molecule. The minimum energy is -98.60 Ha at 1.0 Å. The energy surface and energy difference imply that L-BFGS-B performs not so well since the interatomic distance of 2.5 Å. COBYLA generally shows better accuracy in energy surface, but requires a relatively large evaluation number. The total running times for COBYLA and L-BFGS-B are $1.508 \times 10^4$ and $6.333 \times 10^3$ s, respectively.

Noteworthily, the required qubit numbers for LiH and HF are both equivalent to 12 in the calculation based on the sto3g basis, thus the different running times could be ascribed to their filling fractions of electrons.[37] In LiH, only 1/3 molecular orbitals are occupied, while 5/6 orbitals are occupied in HF. The shorter running time for HF implies that the time cost is larger when the filling fraction is closer to 1/2. Based on the discussion above, we can conclude that the COBYLA performances better in the energy accuracy, but generally requires large evaluation numbers, especially for large molecules. Moreover, the evaluation number of COBYLA is dependent on the interatomic distance, which is generally smaller in the vicinity of the equilibrium distance. L-BFGS-B is more efficient than COBYLA in the optimization process with much smaller evaluation number. However, the accuracy of energy surface from L-BFGS-B is a little lower. Incidentally, although both COBYLA and L-BFGS-B are found to show fascinating efficiency and accuracy in energy surface of $H_4$ previously,[28] their performances could be quite different for complex molecules. The better performance of L-BFGS-B in convergence could be caused by that the gradient gives a definite direction for optimization. The lower energy accuracy in the gradient-based method may be caused by that the optimization enters a local minimum. Recently, a collective optimization algorithm was proposed,[40] which may improve the performance of the gradient-based method in energy accuracy.

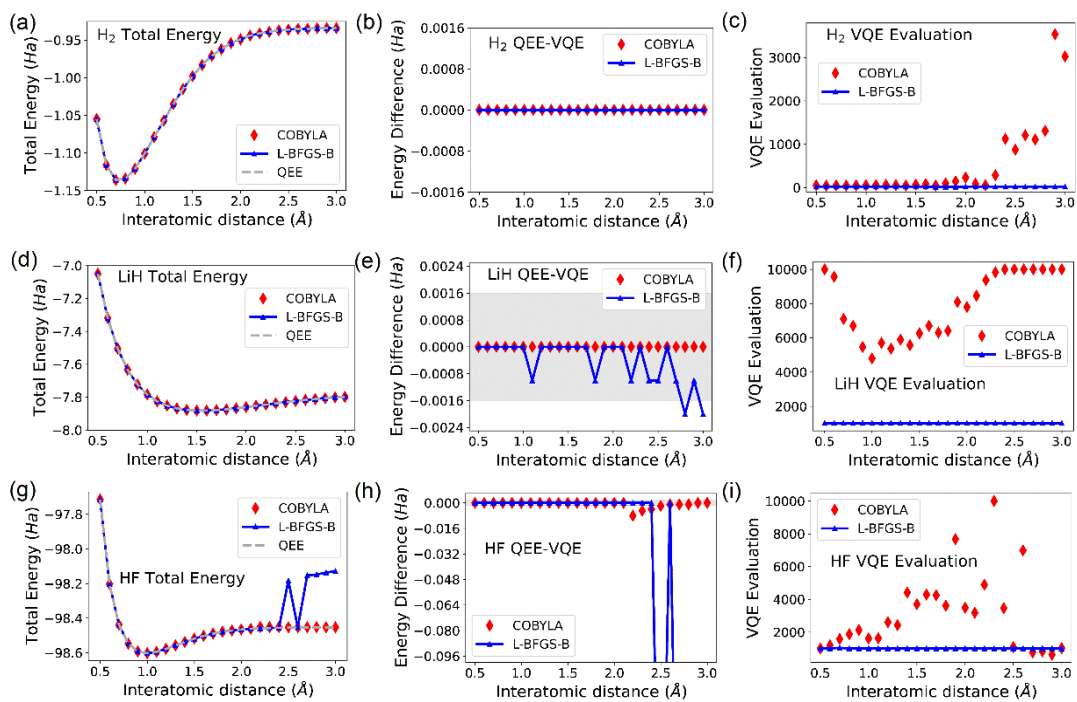

**Figure 1.** (a) The energy surface of $H_2$, the red diamond, blue and dark-grey lines denote the energy surfaces from COBYLA, L-BFGS-B and QEE, respectively. (b) The energy differences for the $H_2$ energy surfaces from COBYLA and L-BFGS-B, compared to that of QEE. (c) The relationship between the interatomic distance and the VQE evaluation number. (d)-(f) are the corresponding energy surfaces, energy differences and evaluation numbers for LiH. (g)-(i) are the corresponding values for HF.

### 4. Basis size influence on the energy accuracy and computation efficiency

In the traditional computational chemistry, basis size plays a vital role in the accuracy of the energy surface. Generally, a larger basis size guarantees a more accurate result, but the corresponding cost also increases significantly. Many efforts are devoted to balance the accuracy and efficiency.[41] In quantum chemistry, most reports adopted the minimal basis size sto3g to show the algorithm.[20, 25] However, the minimal basis size is insufficient for considering the chemistry accuracy. To elaborate the influence of basis size, the energy surfaces for $H_2$ based on different basis sizes are studied and provided in Figure 2. For the VQE calculation, both L-BFGS-B and COBYLA optimization algorithms are also adopted. Figure 2(a) shows the Hartree-Fock energies of $H_2$ based on increasing basis sizes i.e. sto3g, sto6g, minao, 631g, 631g* and 631++g, where minao is defined in the hk.py script in the PySCF package. The corresponding ground state energies from QEE are provided in Figure 2(b). According to the figure, the basis size plays an important role on the energy surface. The larger basis size generally presents a lower energy in the entire range of interatomic distances investigated. The energies from 631g are generally coincide with those from 631g*. Although a few deviations appear between the QEE energies of 631g and 631++g, but the corresponding interatomic distances are far away from its equilibrium bond length. Therefore, it is reasonable to assume that the energy surface of $H_2$ becomes reliable since 631g. The sto3g and sto6g basis present tolerable performances in the range with small interatomic distance, but the corresponding energies deviate significantly with a large interatomic distance. For instance, the energy difference of QEE between those from sto6g and 631g is only 0.005 Ha at the distance of 0.7 Å, but it increases to 0.054 Ha at 3.0 Å. On the contrary, the minao basis shows a reliable performance with a large interatomic distance, but poorly behaves when the interatomic distance is smaller than 1.5 Å. Figure 1(c) shows the energy differences between the QEE and VQE, in which only the COBYLA results are provided for VQE since the energy differences from different basis sizes are both negligible for COBYLA and L-BFGS-B. The tiny energy differences also imply that the single Trotter step in the expansion is sufficient for $H_2$. This behavior is consistent with the previous report which describing Hamiltonian in a particle-hole picture.[42] Figure 1(d) compares the total running time of QEE and those of COBYLA and L-BFGS-B based on different basis sizes. Evidently, the running time increases with increasing size of basis for both QEE and VQE. Moreover, the time of COBYLA is much larger than those from corresponding L-BFGS-B and QEE. For example, the running times of QEE, L-BFGS-B and COBYLA from sto3g are respectively 944, 922 and 1210 s. When the basis size increases to 631++g, the running times increase to 1470, 2992 and 16091 s, respectively. The long running time of COBYLA is mainly caused by the large evaluation numbers as shown in Figure S1 in supporting information (SI). In addition, the required qubits increase with the increasing basis size. For $H_2$, four qubits are employed for sto3g, sto6g and minao, while eight qubits are utilized in 631g and 631g*, and twelve qubits are adopted for 631++g. The basis size influence on the accuracy of energy surface in other molecules are also studied. In Figure S2, the ground state energies of LiH and HF from QEE based on sto3g, sto6g and minao are provided. The larger basis size is not provided here because the required memory is enormous. The required qubit numbers for the molecules investigated are provided in Table S1 in SI. Apparently, the influence of basis size on the accuracy of energy surface is more significant in the molecules with many electrons. Based on the results for the simplest molecule $H_2$, the 631g basis is at least required in considering the chemistry accuracy in quantum computation.

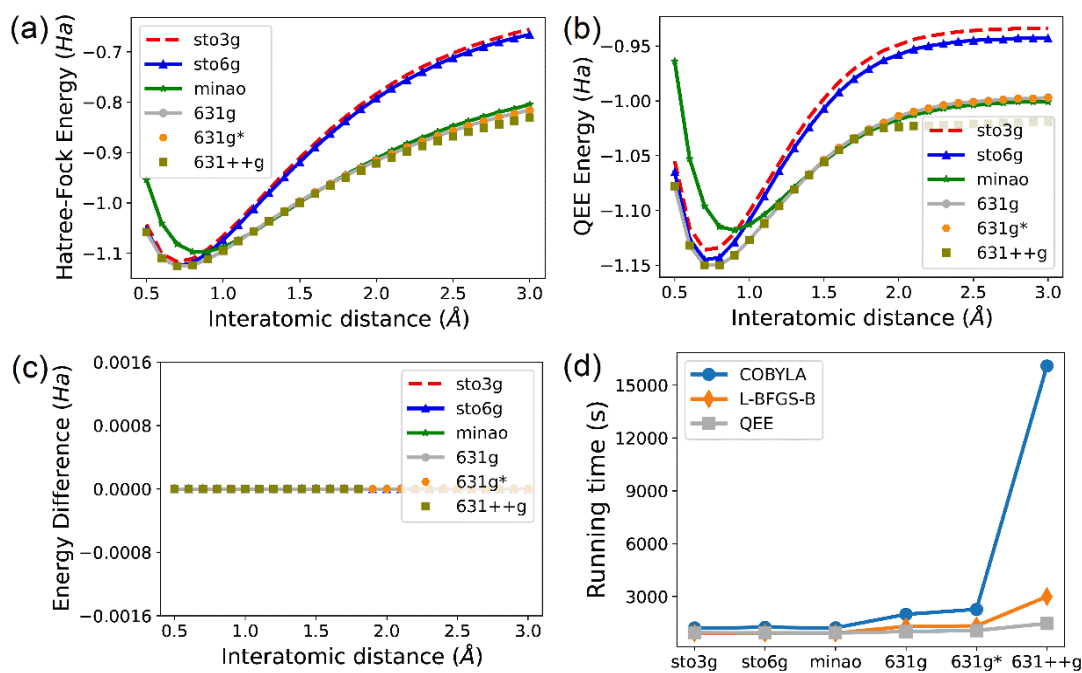

**Figure 2.** (a) The Hartree-Fock energies of $H_2$ based on different basis sizes. The red, blue, green, grey lines, and yellow hexagon and olive square denote the Hartree-Fock energies from the sto3g, sto6g, minao, 631g, 631g* and 631++g basis size, respectively. (b) The ground state energies of $H_2$ from QEE based on the different basis sizes. (c) The energy differences between the QEE and COBYLA for the different basis sizes. (d) The total running times for QEE, L-BFGS-B and COBYLA based on the different basis sizes.

## 5. The CAS approach with different molecular orbitals

Based on the discussion above, the 631g basis is at least required in predicting the energy surfaces of molecules. On the other hand, L-BFGS-B shows much smaller evaluation numbers compared to COBYLA, which is more practical for large molecules. Consequently, the following investigation on the CAS method is mainly based on the L-BFGS-B optimization algorithm and the 631g basis size.

Before investigating the influence of orbital numbers in the performance of CAS,

the superiority of the CAS method is elaborated. With four qubits, the energy surfaces of $H_2$ from the sto3g basis, and that from the CAS method considering the highest occupied molecular orbital (HOMO1) and the lowest unoccupied molecular orbital (LUMO1) based on the 631g basis, are compared in Figure S3 in SI. In addition, the energy surfaces from VQE and QEE based on the 631g basis are also provided as the standard references. Obviously, the energy surface from the CAS method is more accurate than that from the minimal sto3g basis, especially for the values at a large interatomic distance. For instance, at the distance of 2.0 Å, the energy difference between the CAS method and that from QEE of 631g is 0.013 Ha, but the difference between sto3g and 631g increases to 0.065 Ha. Moreover, the evaluation numbers for the CAS method based on 631g, and that from sto3g are also studied in Figure S3(b). The evaluation numbers from the CAS method is close to that with the sto3g basis, and both are much smaller than that from 631g. The total running times for the CAS method and that based on sto3g are respectively 922.7 and 979.5 s. In addition, the running time for the L-BFGS-B method based on 631g is 1303 s. Consequently, with the same number of qubits, the CAS method based on a large basis is more accurate than that with a small basis, at the cost of a little increase in running time. The CAS method is promising for quantum chemistry.

From above, the CAS method with HOMO1 and LUMO1 also shows a deviation from that with corresponding basis. How does the orbital number influence the accuracy and efficiency of VQE based on the CAS approach? In order to address this issue, five molecules ($H_2$, LiH, HF, $N_2$ and $F_2$) are investigated. To facilitate elaboration, the $F_2$ molecular with the largest number of electrons is adopted to elaborate. The results for other four molecules are also provided in Figure S4-S7 in SI. As shown in Figure 3(a), the ground state energy of $F_2$ decreases significantly when the interatomic distance is smaller than 1.6 Å, and the minimum energy is determined to be -198.72 Ha at this distance. After that, the energy shows a slight increase. All the energy curves based on the CAS method with varying occupied and unoccupied molecular orbitals seem to coincide in the entire range of interatomic distances investigated. The red square in Figure 3(a) represents the energy from the CAS method considering HOMO1 and three lowest unoccupied molecular orbitals (LUMO3). The blue diamond shows the value from the CAS method with two highest occupied molecular orbitals (HOMO2) and LUMO3. Similarly, the other symbols represent the values from the CAS method with other different occupied and unoccupied molecular orbitals. Total nine circumstances are studied with both numbers of the occupied and unoccupied orbitals in the range from 1 to 3. To be more explicit, the energy differences of the ground state energies from the CAS method with varying orbitals are provided in Figure 3(b), where the energy surface with the most orbitals i.e. HOMO3 and LUMO3, is adopted as the reference. From the figure, all the values are positive, which implies that the CAS method with HOMO3 and LUMO3 gives the lowest energy in the entire range of interatomic distances investigated. With the same orbital number, more occupied orbitals included in CAS gives a better accuracy of the energy surface. For instance, the values from HOMO3 and LUMO2 are closer to zero, compared to those from HOMO2 and LUMO3. The largest energy difference is determined to be 0.018 Ha at 1.2, 1.3 and 1.4 Å for the former, while the largest energy difference is as large as 0.092 Ha at 1.5 Å for the latter. Similarly, the energy differences from HOMO3 and LUMO1 are generally much smaller than those from HOMO1 and LUMO3. This behavior could be understood that the occupied orbitals outweigh the unoccupied orbitals in the Hamiltonian, based on the Fermi-Dirac distribution. As a result, more occupied orbitals included in CAS could give a more accurate result. Moreover, with different orbitals, more orbitals included in the CAS gives a better result. As an example, with HOMO3, the energy difference decreases with increasing unoccupied orbitals from LUMO1 to LUMO3.

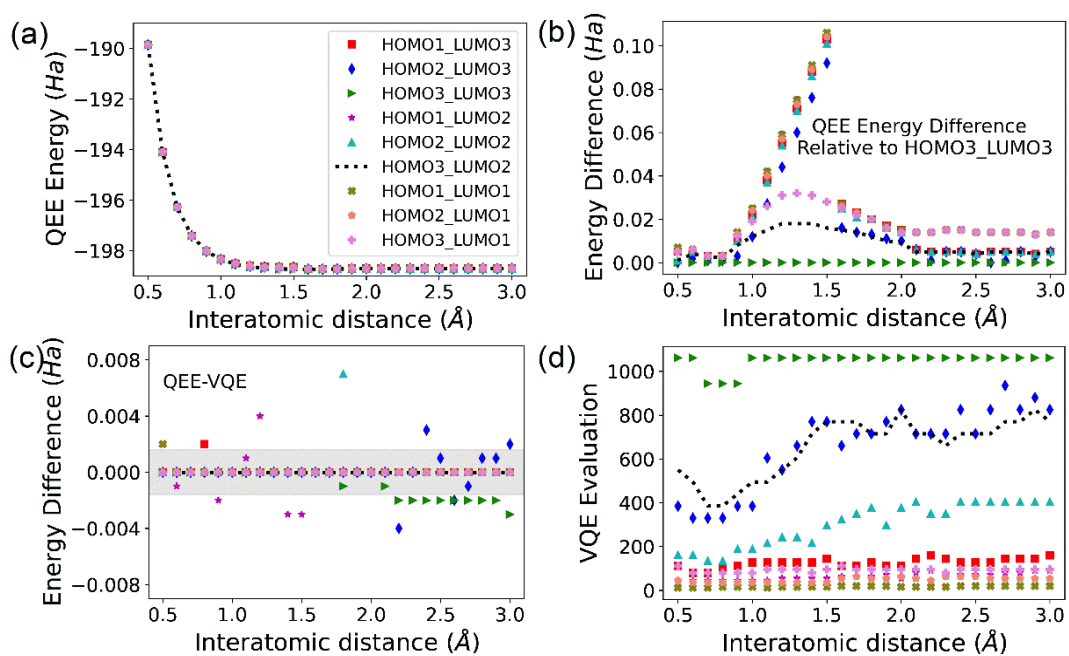

**Figure 3**. (a) The ground state energy of the $F_2$ molecular based on QEE, with various occupied and unoccupied orbitals in CAS from the 631g basis. The red square denotes the value calculated with HOMO1 and LUMO3. The blue diamond represents the data considering HOMO2 and LUMO3. The green right triangle shows the value calculated with HOMO3 and LUMO3. The magenta asterisk is the value considering HOMO1 and LUMO2. The cyan upper triangle presents the value considering HOMO2 and LUMO2. The black crisscross represents the value considering HOMO3 and LUMO2. The olive cross shows the value considering HOMO1 and LUMO1. The salmon pentagon and violet bold crisscross respectively represent the values considering HOMO2 and LUMO1, and HOMO3 and LUMO1. (b) The energy differences between the values calculated with varying orbitals, relative to the energy calculated with HOMO3 and LUMO3. (c) The energy difference between those from QEE and VQE, with the same

molecular orbitals in the CAS. (d) The evaluation numbers for L-BFGS-B based on various orbitals in CAS.

Figure 3(c) shows the energy difference between the QEE and corresponding VQE. With the same number of orbitals, the more occupied orbitals include, the smaller energy difference generally appears. As an example, the energy difference from HOMO3 and LUMO2 is smaller than that from HOMO2 and LUMO3. Figure 3(d) presents the evaluation numbers during VQE optimization. The evaluation number increases with the increasing orbitals included in CAS. With HOMO3, the average evaluation numbers are respectively 92.92, 655.8 and 1048 with the number of unoccupied orbitals from 1 to 3. Similarly, with LUMO3, the average evaluation numbers are respectively 125.5, 660.0 and 1048 with the number of occupied orbitals from 1 to 3. From the data, with the same number of orbitals, LUMO requires a larger evaluation number during the optimization process. In addition, the evaluation number also correlates with the filling fraction of orbitals. Taking four molecular orbitals in CAS as an example, the evaluation number of the CAS method with HOMO2 and LUMO2 is larger than that with HOMO1 and LUMO3, and as well as that with HOMO3 and LUMO1. The influence of filling fraction of orbitals on VQE evaluation is consistent with the previous discussion for comparing the running times of the LiH and HF molecules. Based on the discussion above, it is reasonable to include more occupied orbitals in CAS based on the limited qubits in describing the molecular orbitals. With the same number of qubits, the more occupied orbitals include, the better energy accuracy and smaller evaluation number appear for the VQE calculation. Moreover, the optimization process generally requires more evaluation numbers when the filling fraction closer to 1/2. The results for other four molecules shown in Figure S4-S7 generally consist with the findings from the $F_2$ molecule. Based on their different electronic structures, only HOMO1 exists for the occupied orbitals of $H_2$, and HOMO1 and HOMO2 occur for the occupied orbitals of LiH.

In order to understand the influence of the molecular orbital on the energy accuracy and calculation efficiency, the detailed Hamiltonian $H$ and corresponding eigenvector $|\Psi(\theta)>$ based on different orbitals are further studied. At the equilibrium distance of 1.6 Å, the Hamiltonian and the eigenvector of the $F_2$ molecule with HOMO1 and LUMO3, HOMO2 and LUMO2, and HOMO3 and LUMO1 are respectively examined, corresponding results are provided as supplemental files. For instance, the F2_H_HOMO1_LUMO3 and F2_Eig_HOMO1_LUMO3 files are respectively the Hamiltonian and optimized eigenvector of the $F_2$ molecule with HOMO1 and LUMO3. From these files, the Hamiltonian with HOMO1 and LUMO3, and HOMO3 and LUMO1 are both summed by 61 Hamiltonian terms, such as IIIIIIII and IIIIIIIZ, with I and Z denoted the Pauli matrix. Correspondingly, the Hamiltonian with HOMO2 and LUMO2 has 97 terms. Regarding to the eigenvectors, the one with HOMO1 and LUMO3 is prepared by 1393 quantum gates, and that with HOMO3 and LUMO1 is build by 1445 quantum gates. Moreover, the eigenvector with HOMO2 and LUMO2 is made by 2656 quantum gates. Evidently, the Hamiltonian and eigenvector of HOMO2 and LUMO2 possess more Hamiltonian terms and quantum gates than the ones with HOMO1 and LUMO3, or with HOMO3 and LUMO1. The different numbers in Hamiltonian terms and quantum gates are arose from the different possible excitation modes. According to the second quantization description of Hamiltonian $H = \sum_{pq} h_{pr} a_p^+ a_r + \frac{1}{2} \sum_{pqrs} h_{pqrs} a_p^+ a_q^+ a_r a_s$ ( $h_{pr}$ and $h_{pqrs}$ are the one and two electron integrals), and the detailed formulation of UCCSD as described in the computational details, the CAS with HOMO2 and LUMO2 generally possess more possible excitation modes.

Based on the more adjust parameters in the eigenvector, the evaluation number in VQE is larger when the filling fraction of orbitals is closer to 1/2.

### 6. The impact of electronic structure

After the influences of the orbitals in CAS on the energy accuracy and efficiency of quantum calculation are studied, the impact of electronic structure in a molecule on VQE are additionally studied. These five molecules studied have different occupied orbitals. To remove the influence of different orbitals, the energy differences between QEE and VQE, and evaluation numbers for optimization processes are compared based on the CAS method with only HOMO1 and LUMO1. As shown in Figure 4(a), the energy differences between the QEE and VQE are generally below the chemical accuracy for $H_2$ and $F_2$. Regarding to the LiH and HF molecules, the energy differences are below the chemical accuracy when the interatomic distances are smaller than 2.0 Å. On the contrary, the energy difference for $N_2$ is relatively large since 1.5 Å. The value with the largest magnitude is determined to be -0.064 Ha at 2.9 Å. For the evaluation numbers shown in Figure 4(b), the values for $H_2$ and $F_2$ are smaller than those in other three molecules. The averaged evaluation numbers for $H_2$, LiH, HF, $N_2$ and $F_2$ are 19.23, 24.15, 22.15, 26.76 and 17.07. The corresponding running times of VQE are respectively 979.5, 1251, 1430, 1440 and 1401 s. Thus, the cost time for each evaluation is 50.93, 51.80, 64.56, 53.81 and 82.07 s, respectively for $H_2$, LiH, HF, $N_2$ and $F_2$. Apparently, the cost time for each evaluation increases with the increasing maximum nuclear charge in a molecule. The increase in the cost time of each evaluation may relate to the eigenvector preparation process, since the eigenvector is proportional to the maximum nuclear charge.[37] In a word, VQE shows better performances in $H_2$ and $F_2$, compared to that in LiH and HF. This behavior implies that a high energy accuracy and a small evaluation number for VQE are prone to appear in the molecule with homogeneous elements. However, the energy difference between QEE and VQE, and the evaluation number are both much larger in $N_2$. This performance may relate to the large bonding energy in $N_2$. As is well-known, only a single bond is formed in the $H_2$, LiH, HF and $F_2$ molecules, but a triple one is bonded in $N_2$. As an estimate, the bonding energy could be simplified as the energy difference between the energy at the largest interatomic distance of 3.0 Å and the minimum value in the energy surface. According to the energy surface, the bonding energy in $N_2$ is as high as 0.597 Ha, while the values in $H_2$, LiH, HF and $F_2$ are respectively 0.147, 0.045, 0.087 and 0.022 Ha. To form the strong triple bond in $N_2$, the VQE generally requires more evaluation to obtain a stable electron distribution, and the predicted energy surface shows a larger error.

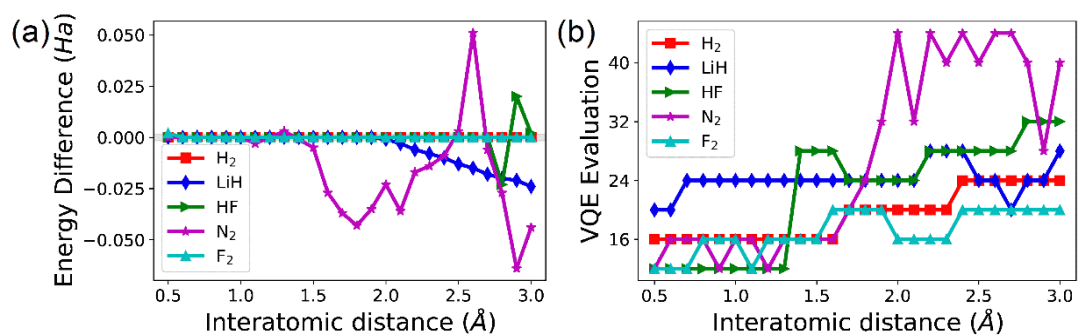

**Figure 4**. (a) The energy differences between the values calculated from QEE and VQE based on the CAS method with HOMO1 and LUMO1. The red square, blue diamond, green right triangle, magenta asterisk and cyan upper triangle respectively represent the energy difference for the $H_2$, LiH, HF, $N_2$ and $F_2$ molecules. (b) The evaluation numbers for the VQE optimization based on the L-BFGS-B algorithm.

## 7. Conclusion

In summary, we have studied the influences of optimization method and basis size on the accuracy of energy surface and efficiency of quantum simulation based on the VQE method. Different from the previous report, the performances of L-BFGS-B and COBYLA are quite different, especially for large molecules. L-BFGS-B is more practical for VQE since it converges more quickly than COBYLA, although the accuracy of corresponding energy surface is a little lower. The basis size plays a significant role in the energy surface. The 631g basis is at least required based on the energy surfaces of $H_2$. For practical applications, the CAS method are considered. For a specific molecule, the CAS method with more orbitals included gives a better energy accuracy, but the corresponding computing cost increases. With the same number of orbitals, more occupied orbitals considered can enhance the accuracy of the energy surface, and the corresponding optimization process converges more quickly. Moreover, the filling fraction of orbitals also influences the VQE optimization process, where the optimization process generally requires more evaluation numbers when the filling fraction is closer to 1/2. In addition, the bonding energy and the maximum nuclear charge in a molecule also influences the accuracy and efficiency of VQE, where VQE performs better in a molecule with a weaker bond strength. We hope our work could provide a guidance for the practical application of VQE in quantum computers.

**Notes**

The authors declare no competing financial interest.

**Acknowledgement**

The authors acknowledge the financial support from National Natural Science Foundation of China (No. 11974252).21

and optimized wave-function expansions. *Phys. Rev. A* **2018,** *98*, 022322.

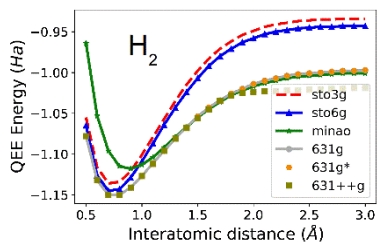
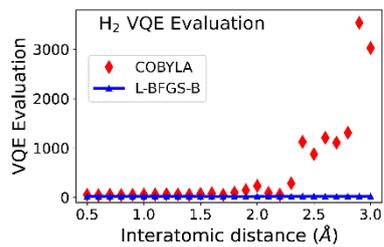
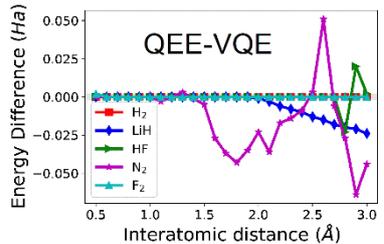

**Table of Contents**